\documentstyle[prl,aps,multicol,epsf]{revtex}

\makeatletter
\ifcase\@ptsize
  \font\tenmsy=msbm10
  \font\sevenmsy=msbm7
  \font\fivemsy=msbm5
  \font\tenmsx=msam10
  \font\sevenmsx=msam7
  \font\fivemsx=msbm5
\or
  \font\tenmsy=msbm10 scaled \magstephalf
  \font\sevenmsy=msbm8
  \font\fivemsy=msbm6
  \font\tenmsx=msam10 scaled \magstephalf
  \font\sevenmsx=msam8
  \font\fivemsx=msam6
\or
  \font\tenmsy=msbm10 scaled \magstep1
  \font\sevenmsy=msbm8
  \font\fivemsy=msbm6
  \font\tenmsx=msam10 scaled \magstep1
  \font\sevenmsx=msam8
  \font\fivemsx=msam6
\fi
\newfam\msyfam
\textfont\msyfam=\tenmsy
\scriptfont\msyfam=\sevenmsy
\scriptscriptfont\msyfam=\fivemsy
\newfam\msxfam
\textfont\msxfam=\tenmsx
\scriptfont\msxfam=\sevenmsx
\scriptscriptfont\msxfam=\fivemsx
\def\Bbb{\ifmmode\let\next\Bbb@\else
\def\next{\errmessage{Use \string\Bbb\space only in math mode}}\fi\next}
\def\Bbb@#1{{\Bbb@@{#1}}}
\def\Bbb@@#1{\fam\msyfam#1}
\def\Aaa{\ifmmode\let\next\Aaa@\else
\def\next{\errmessage{Use \string\Aaa\space only in math mode}}\fi\next}
\def\Aaa@#1{{\Aaa@@{#1}}}
\def\Aaa@@#1{\fam\msxfam#1}
\newfam\euffam
\font\sixeuf=eufm6
\font\eighteuf=eufm8
\font\twelveeuf=eufm10 scaled\magstep1
\textfont\euffam=\twelveeuf
\scriptfont\euffam=\eighteuf
\scriptscriptfont\euffam=\sixeuf

\makeatother

\newcommand{\BZ}{{\Bbb{Z}}}

\def\vec#1{{\rm\bf#1}}
\def\vac#1{|{#1}\rangle}

\def\vev#1{\langle{#1}\rangle}

\def\Vvev#1{\left\langle\kern-5pt\left\langle{%
  #1}\right\rangle\kern-5pt\right\rangle}
\def\Vvevs#1{\left\langle\kern-4pt\left\langle{%
  #1}\right\rangle\kern-4pt\right\rangle}
\def\Vvevss#1{\left\langle\kern-3pt\left\langle{%
  #1}\right\rangle\kern-3pt\right\rangle}

\def\del{\partial}

\def\jb{\bar{\jmath}}

\def\be{\begin{equation}}
\def\ee{\end{equation}}
\def\ba{\begin{array}}
\def\ea{\end{array}}
\def\bea{\begin{eqnarray}}
\def\eea{\end{eqnarray}}
\def\bean{\begin{eqnarray*}}
\def\eean{\end{eqnarray*}}
\def\bl{\begin{list}{}{}}

\begin{document}
\draft

\title {Fun with Quantum Dots}
\author{Mic$\hbar$ael Flohr\footnote{
\begin{tabular*}{6.9in}[t]{rl@{\extracolsep{\fill}}r}
    homepage: & {\tt http://www.mth.kcl.ac.uk/\~{}flohr/} & KCL-MTH-98-56\\
    email:    & {\tt flohr@mth.kcl.ac.uk}                 & cond-mat/9811288
  \end{tabular*}
}\footnote{Supported by EU TMR network fellowship no.\ FMRX-CT96-0012}}
\address{Department of Mathematics, King's College London\\ 
The Strand, London WC2R 2LS, United Kingdom}
\date{\today}
\maketitle

\begin{abstract}
We consider quantum dots with a parabolic confining potential.
The qualitative features of such mesoscopic systems as functions
of the total number of electrons $N$ and their total angular
momentum $J$, e.g.\ magic numbers, overall symmetries etc., are
derived solely from combinatoric principles.
The key is one simple hypothesis about
such quantum dots yielding a basis of states (different from the
usual single electron states one starts with) which is extremely
easy to handle. Within this basis all qualitative features are
already present without the need of any perturbation theory.
\end{abstract}


\begin{multicols}{2}
\narrowtext

\section{Introduction}

Quantum dots with a parabolically confined potential are a particularly
simple species of a phenomenon enjoying increasing interest \cite{qdotgen}. 
The basic novelty with such systems is
their being {\em mesoscopic}, i.e.\ the fact that they live on the
edge between classical and quantum physics. Quantum dots are sometimes
called artificial atoms, which illuminates their mesoscopic nature.
Indeed, although much larger than a real atom, the angular momentum $j$
of the electrons is quantized, and hence, via the classical relation
$j\propto r^2$, its orbital radius.

It has been known for some time \cite{magic,smagic} that the total potential
energy of a (parabolic confining) quantum dot with $N$ electrons exhibits
local minima for certain amounts of the total angular momentum $J$, called
magic numbers. In fact, the local minima occur when 
\be\label{eq:magic}
  J(N,k) = \frac{N(N-1)}{2} + kN\,,\ \ \ \ k\in\BZ_+\,.
\ee
The quantum nature of the system becomes even more apparent when we
compare the behaviour for $N\leq 5$ electrons with the one for $N\geq 6$
electrons. Classically, we expect that the lowest energy configuration
for $N\leq 5$ electrons is a regular $N$-gon, while for $N\geq 6$ 
electrons it should be a regular $(N-1)$-gon with one electron in the
center. However, whenever $k=(N-1)k'$ with $k'\in\BZ_+$, the configuration
at such special magic numbers is much less symmetric than the classical
counterpart. In fact, it behaves much more like a Laughlin-type
quantum droplet.

We will demonstrate in this letter that the appearance of magic numbers
as well as the loss of symmetry for the special magic numbers satisfying
\be\label{eq:smagic}
  J(N,k'(N-1)) = N(N-1)\left[k'+ \frac{1}{2}\right]\,,\ \ \ \ k'\in\BZ_+\,,
\ee
can be explained by pure combinatorics. To do so, we need a bit of
preparation.

We assume that the external magnetic field is strong enough to completely
polarize all electrons. Then, only one electron can occupy an eigenstate
of the angular momentum operator with eigenvalue $j\in\BZ_+$. The system
characterized by the total number of electrons $N$ and the total angular
momentum $J$ may now reside in any quantum mechanical state of the form
$[j_1,\ldots j_N]$ such that $0\leq j_1<j_2<\ldots<j_N$ and
$\sum_{i=1}^N j_i = J$, where the $j_i$ denote the angular orbitals in
which the electrons may stay. Of course, the $N$ electrons are
identical particles meaning that each angular momentum eigenstate is
a one-particle state antisymmetrized over all the electrons. To count the
number of possible configurations one simply has to expand the combinatoric
partition function
\be
  Z(y,q) = \prod_{n=0}^{\infty}(1+yq^n) = \sum_{N,J}c_{N,J}y^Nq^J\,.
\ee
The product includes the term $n=0$
since one electron may reside in the $j=0$ state. 
The reader might convince themselves that $c_{N,J}$ can be quite large even
for very small electron numbers and medium sized total angular momenta.
The true state is a complicated superposition of these configurations
which depends on the precise interaction of the electrons giving each
configuration a different weight. The problem is that each configuration
is a sum of $N$ single-particle states. The latter are eigenstates 
only for the free Hamiltonian without interactions. At this point,
numerical methods are usually employed to diagonalize the physical
Hamiltonian which includes the Coulomb interaction as well as other
effects \cite{numeric}.

Here, we will take a different approach. First, we note that any
configuration $[j_1,\ldots,j_N]$ might contain so-called {\em blocks\/}
of adjacent angular orbitals which are all occupied. Such a block
$j_i,j_i+1,\ldots,j_i+n-1$ is uniquely characterized by two numbers, 
namely the number $n$ of adjacent occupied orbitals, and the number $\jb=
(j_i+j_i+1+\ldots+j_i+n-1)/n=j_i+\frac12(n-1)$. It is clear that we can
enumerate all the configurations also by the blocks they decompose into,
$[(n_1,\jb_1),(n_2,\jb_2),\ldots,(n_{\ell},\jb_{\ell})]$. 
Note that $\jb$ might be half-integer, and that $(1,\jb)$ denotes 
an orbital which does not have an occupied neighbouring orbital. Of course, 
$1\leq \ell\leq N$ with the extreme cases given by either one complete
block or only isolated single orbitals. Since the angular momentum orbital
also defines a radial quantization, electrons in a block are also
radially adjacent. In some abuse of language we might view each such
configuration, projected on the radial component, as a configuration 
of a spin chain where spins can either be occupied or not. This leads us
to the one hypothesis we will make in this paper:

{\sc Hypothesis:} {\em The electrons within a block $(n,\jb)$ are 
(sufficiently strongly) 
coupled due to spin-spin and spin-orbit interactions to form a composite 
state of charge
$n$ and ``spin'' $\jb$, while there is no (or negligible) spin-spin and
spin-orbit
interaction among different blocks.}

Clearly, this hypothesis is the equivalent of a nearest-neighbour
interaction approximation. The clue is that we can treat a block as
a collective state. In particular, antisymmetrizing is performed seperately
for each block, which will influence certain multiplicities. Also, the
Coulomb interaction between two blocks $(n_1,\jb_1)$ and $(n_2,\jb_2)$
is calculated classicaly as given by the potential between two rings of
charge $n_1$ and $n_2$ with radii $\sqrt{\jb_1}$ and $\sqrt{\jb_2}$
respectively. Here and in the following we adopt a system of units in which
all proportionality constants are put to one, since we are only concerned
with qualitative features. However, in most computations only ratios
appear, making them independent of our particular choice of units. Our choice
is motivated by keeping everything as close as possible to pure integers
emphasizing the combinatorical nature of our approach.

Our basis of composite states does already account for spin-spin
effects, the remaining other important effect being the Coulomb
interaction. However, to further simplify all calculations, we make
the following observation: It has been shown numerically
that the qualitative behavior of classical dots is quite independent of the
precise nature of the repulsive interaction $V(\vec{r}_i,\vec{r}_j)=
e_ie_j|\vec{r}_i-\vec{r}_j|^{-\gamma}$, with $\gamma=1$ yielding the
Coulomb case, see e.g.\ \cite{DMV98}. The same should hold for quantum
dots, as our results confirm a posteriori for $\gamma=2$.
This is called universality, a property which quantum dots
share with Laughlins wave functions for the fractional quantum Hall
effect at filling factors $\nu=1/(2p+1)$. However, the formul\ae\ for
the potential energy between blocks and, in particular, the self-energy
of a block, can only be given in a simple closed form in $n$ and $\jb$ for
$\gamma$ even. Since the spirit of this paper is simplicity, we will
therefore put $\gamma=2$ with the silent understanding of universality.
We have checked for a wide range of $N,J$ that the qualitative features 
are indeed not affected by this.

Thus, within our setting, the self-energy of a block $(n,\jb)$ is
given, in the equilibrium state, by the potential energy of a regular
$n$-gon of radius $\sqrt{\jb}$, which yields
\be\label{eq:En-gon}
  E_{(n,\jb)} = \frac{(n+1)n(n-1)}{3!\,4\jb}\,.
\ee
Similarly, the potential energy between two such blocks is simply
\be\label{eq:E2n-gons}
  E_{(n_1,\jb_1),(n_2,\jb_2)} = \frac{n_1n_2}{|\jb_1 - \jb_2|}\,.
\ee
Note that the squared radii in these formul\ae\ have been replaced by
the ``spins'' of the composite states, bringing everything down to
rational numbers. It will be imporant later that, if we were to
include the precise proportionality factor of the relation 
$j = m_e\omega r^2$, it would yield a common factor to all these
partial energies. We are now ready to proceed in finding magic numbers.

\section{Magic Numbers}

All we have to do is calculate the energies for all block configurations
according to the above formul\ae, where we neglect any pure quantum 
mechanical effects (such as the rest energy $\frac12 N\hbar$) as well as
the kinetic term (which is linear in $J$) yielding the so-called excitation
energy
\bea\label{eq:Etot}
  E_m&\equiv& E[(n_1,\jb_1),\ldots,(n_{\ell},\jb_{\ell})]\nonumber\\
  &=&
  \sum_{i=1}^\ell E_{(n_i,\jb_i)} +
  \sum_{i<k} E_{(n_i,\jb_i),(n_k,\jb_k)}\,,
\eea
where we have assumed that we have enumerated all configurations (in a
completely arbitrary manner) by $m$. We will now assume that the normalized
ground state of the quantum dot can be written as a linear
combination $\vac{\Psi} = \sum_m S_m
\exp(-\frac12\beta E_m)\vac{\Psi_m}$, with the $\vac{\Psi_m}$ forming an
ortho-normal system, 
i.e.\ each configuration $m$ will be
weighted accordingly by its Boltzman factor $\exp(-\beta E_m)$. The
inverse temperature $\beta$ is as yet unknown, but can easily be
determined by solving the normalization condition
\be\label{eq:norm}
  {\cal Z}(\beta) = \sum_m S_m\exp(-\beta E_m) = 1\,.
\ee
Please note that this is much easier than it looks! Putting $x=\exp(-\beta)$,
one only has to find the unique real zero of the polynomial 
${\cal Z}(x')-1$ which lies in the interval $x'\in[0,1]$. 
Here $x'=x^{1/M}$ with $M$ the least common multiple of all denominators 
of the rational(!) numbers $E_m$, and the polynomial 
${\cal Z}(x')-1\in\BZ[x']$. Finding
this one zero can be done to arbitrary percision and very quickly, even for
huge polynomials. The $S_m$ are symmetry factors which take into
account that each block is antisymmetrized individually, but also that
each block $(n,\jb)$ has a discrete $\BZ_n$ symmetry. Hence individual
antisymmetrization yields for each such block a symmetry factor $n!|\BZ_n|^{-1}
=(n-1)!$. This has to be multiplied by the number of ways to distribute
the $N$ electrons among the $\ell$ blocks of a given configuration. Hence
\be
  S_m\equiv S[(n_1,\jb_1),\ldots,(n_{\ell},\jb_{\ell})] =
  {N\choose \ell}\prod_{i=1}^{\ell}(n_i-1)!\,,
\ee
such that the probability of the $m$-th configuration is 
$p_m=S_m\exp(-\beta E_m)$.

Alternatively, we could leave $\beta$ undetermined, treating it as an outer
parameter of the system. This amounts in allowing different values for the
normalization constant ${\cal Z}(\beta)$. Again, we have checked that
the qualitative features vary weakly with $\beta$. Fixing $\beta$ as above
yields -- in a loose sense -- the quantum mechanical (inverse) temperature,
a measure of how likely the system fluctuates\linebreak
\begin{figure}[h]
    \begin{center}
    \epsfxsize=85mm 
    \epsfysize=80mm 
    \leavevmode
    \epsfbox{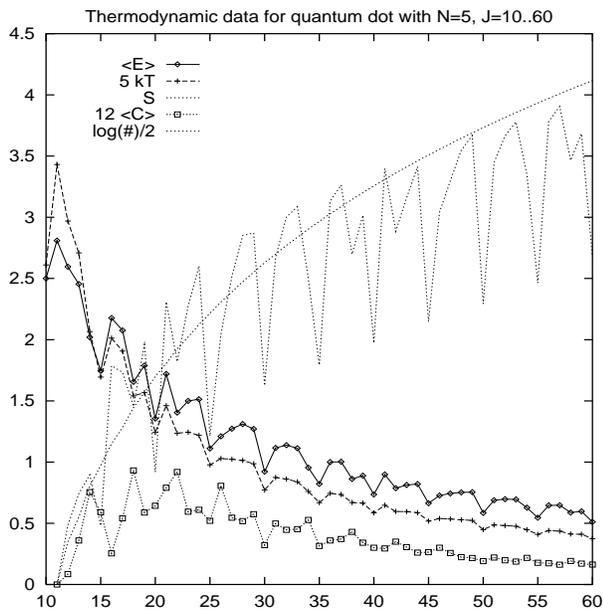} 
    \end{center}
    \caption{Thermodynamic data for a parabolic confining quantum dot with
    $N=5$ electrons and total angular momentum $10\leq J \leq 60$. The
    magic numbers at which the energy as well as the temperature exhibit
    local minima can clearly be seen. At these numbers, the entropy 
    also has local minima indicating that the system has higher order
    for these angular momenta. Indeed, the system is closest to a cristalline
    structure at the magic numbers. The behaviour of the entropy is contrasted
    with the logarithm of the number of configurations.}
\label{fig:T5}
\end{figure} 
\noindent quantum mechanically between its different
states (neglecting thermal fluctuations due to a heat bath).

\begin{figure}[h]
    \begin{center}
    \epsfxsize=85mm 
    \epsfysize=80mm 
    \leavevmode
    \epsfbox{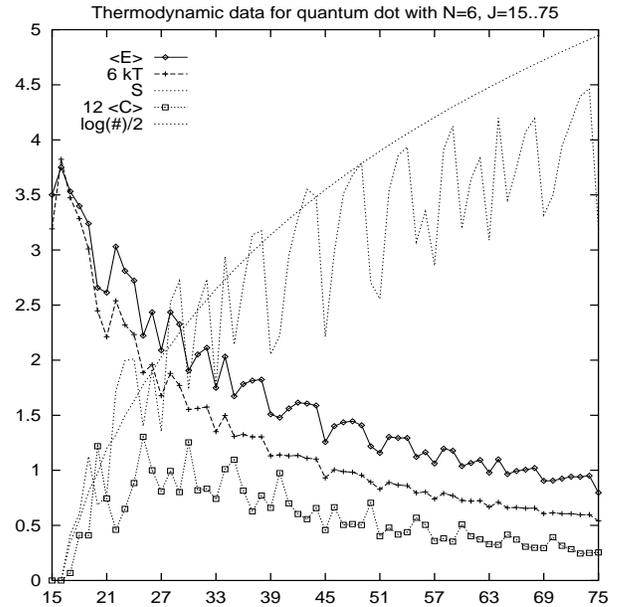} 
    \end{center}
    \caption{Thermodynamic data for a parabolic confining quantum dot with
    $N=6$ electrons and total angular momentum $15\leq J \leq 75$. Compare
    the magic numbers with the ones for $N=5$. A close look reveals that
    in addition to the period $N=6$ of magic numbers, equation 
    (\ref{eq:magic}), another period of $N-1=5$ is superimposed according
    to [3] which coincides with the former for $J=45$. 
    Of course, the same is true for the $N=5$ quantum dot, but less prominently 
    expressed. Note that magic numbers can already be
    seen for very small $J\sim J_{{\rm min}}=N(N-1)/2$.}
\label{fig:T6}
\end{figure}

After determining $\beta$, we have the complete partition function of
our quantum dot system which we now treat as a system in statistical
mechanics. Therefore we can easily determine observables such as the
energy $\vev{E} = \del_{\beta}\log({\cal Z})=\sum_m p_m E_m$, the specific 
heat $\vev{C}\equiv\beta^{-2}C_v=\del_{\beta}^2\log({\cal Z})=
\vev{E^2}-\vev{E}^2$,
and the entropy $S=-\sum_m p_m\log(p_m)$. 
As figures \ref{fig:T5} and \ref{fig:T6} demonstrate,
the magic numbers show up in these observables exactly as they are
supposed to do. We compare the most interesting cases, $N=5$ and $N=6$.
The temperature $1/\beta$ and the specific heat are multiplied by
suitable factors to improve their appearance in the plots. It should be
noted that all these calculations, whose results are collected 
in the plots, only take a few minutes with a small and simple C-program on 
a typical workstation. 

Naturally, the ``smoothness'' of the curves shown 
in figures \ref{fig:T5} and \ref{fig:T6} 
increases with $N$ and for each $N$ with $J$ due to the fast growth of
$c_{N,J}$. However, we find it noteworthy that a mesoscopic system of only
2 or 3 particles already shows the same features, i.e.\ can
successfully be approximated by a statistical system. One can
interpret this fact that quantum mechanics in a mesoscopic system shows 
itself mainly by moving the sys-\linebreak
\noindent tem from a purely classical configuration to a statistical
one. To show more clearly what we mean by this, we will consider the
radial probability distribution, i.e.\ $p(R)_{N,J} = \int_0^{2\pi}
|\Psi(R,\phi)_{N,J}|^2\,{\rm d}\phi$ for given $N,J$. The point is that
we do not even have to calculate anything of the wavefunction 
$\Psi(R,\phi)_{N,J}$. Since angular momentum and orbital radius are
interlinked, we only need to consider the formal sum
\be
  \tilde R(q) = \sum_m p_m\sum_{i=1}^{\ell_m}\frac{n_i}{N}q^{\jb_i}
       = \sum_{j=0}^J \tilde p_jq^j\,,
\ee
where again $m$ stands for the $m$-th configuration of
blocks $[(n_1,\jb_1),\ldots,(n_{\ell},\jb_{\ell_m})]$. 
The mathematically inclined
reader may wish to consider this as some kind of discrete Fourier 
transform of the radial or, more precisely, angular momentum probability
distribution. The radial distribution can then be read off as 
$p(R=\sqrt{j}) = \tilde p_j$. However, we assumed here that the composite
block state $(n,\jb)$ is localised at $\jb$. Since this state is a
collective state out of $n$ electrons occupying the orbits from
$\jb-\frac{n-1}{2}$ to $\jb+\frac{n-1}{2}$, it will certainly be smeared out
over this region in angular momentum space. In the spirit of treating the
system as a statistical mechanics ensemble, the most natural distribution
of the composite state is a binominal distribution centered at $\jb$ with
width $\frac{n}{2}$. One could equally take the point of view that a
Gaussian distribution is more natural for a quantum mechanical state. 
However, since we are only interested in a qualitative analysis, we may
neglect the difference between these two distributions (one may
convince onself that both distributions lead to very similar functions
$p(r)$).
Since $\jb$ may be half integer, the correct binominal distribution yields
\bea\label{eq:radial}
  \hat R(q) &=& \sum_m p_m\sum_{i=1}^{\ell_m}\frac{n_i}{N}
         \sum_{l_i=0}^{2n_i-2}{2n_i-2\choose l_i}
         \frac{q^{\jb_i+\frac{1}{2}(l_i+1-n_i)}}{2^{2n-2}}\nonumber\\
       &=& \sum_{j=0}^J \hat p_jq^j
\eea
as the radial distribution function. We are now ready to discuss the issue
of symmetry loss, which is observed at the special magic numbers eq.\ 
(\ref{eq:smagic}) for $N\geq 6$.

\section{Symmetry Loss}

The smallest non-trivial special magic number for $N=5$ is $J(5,4)=30$, and for
$N=6$ it is $J(6,5)=45$. Direct numerical analysis shows that around these 
numbers, $(N-1)$-gon symmetry with an occupied center is the preferred
symmetry, whereas at the precise special magic numbers, and $N\geq 6$, 
symmetry is even more weakly expressed. One speaks of a fluid-like state at
these numbers (in analogy to Laughlins quantum droplets), and says that the 
quantum dot with these particular values of total angular momentum
$J=(N,k'(N-1))$ has filling factor 
\be\label{eq:nu}
  \nu = \frac12 \frac{N(N-1)}{J} = \frac{\frac12 N(N-1)}{N(N-1)(k'+\frac12)}
       = \frac{1}{2k'+1}\,.
\ee
Before studying the radial distributions, we would like to explain this fact, 
and also that it is only observed for $N>5$. For smaller particle numbers, 
$\BZ_N$ symmetry is not broken (or only very slightly). Our considerations 
are made from very simple general assumptions and do not rely on any
numerical studies.
The key ingredient is the following relation between different magic numbers:
\bea\label{eq:rec}
  \lefteqn{J(N,(N-1)k') =}\\
  && J(L,(L-1)k') + J(N-L,(N+L-1)k' + L)\,.\nonumber
\eea
Note that this relation can be applied recursively! The meaning is that for 
the special magic numbers, there are several ways to build a state out of 
smaller magic units. However, the above relation is quite different from the
trivial relation $J(N,k) = J(L,k) + J(N-L,k+L)$, which just divides one 
block of angular momentum eigenstates into two adjacent ones. The former
relation is to be understood as follows. If the offset $k$ is divisible by 
$N-1$ and sufficient large, $\BZ_N$ symmetry can be replaced by
$\BZ_L$ symmetry (with a rather small offset) within $\BZ_{N-L}$ symmetry 
(with an even larger offset). This is possible, because this distribution 
of angular momenta among the particles gives the same magic number, but with 
two sufficiently separated blocks to allow separate antisymmetrization. 

Clearly, this alone does not force loss of symmetry. The question is, which 
of these possible distributions has the lowest energy, and how close do the 
energies of the other distributions come to the minimal one. To answer these
questions, one only needs the formul\ae\ (\ref{eq:En-gon}), 
(\ref{eq:E2n-gons}), and (\ref{eq:Etot}), and compare energies of
different configurations. If the system were classical, we would only have to
compare $E[(N,J/N)]$ with $E[(N-1,J/(N-1)),(1,0)]$. The former energy
is smaller than the latter as long as $N<\frac14(13+\sqrt{73})<6$. Since
a quantum dot is quantum mechanical by definition, all possible
configurations contribute and we have to calculate
the full partition function ${\cal Z}$ instead.

The above formula (\ref{eq:rec}) is recursive. Moreover, one of the magic 
numbers on the right hand side is again a special magic number.
Thus, it appears that the fluid-like quantum state is due to the fact 
that (for ``large'' $N$) there are a lot of different antisymmetrizations
possible at these numbers which -- as we are going to show -- have all 
comparable similar energies, where the energy of the center occupied 
$(N-1)$-gon configuration is the minimal one for $N>5$. 
There even exist antisymmetrizations with three or more nested blocks,
whenever $N$ is large enough. 

\begin{figure}[h]
    \begin{center}
    \epsfxsize=85mm 
    \epsfysize=80mm 
    \leavevmode
    \epsfbox{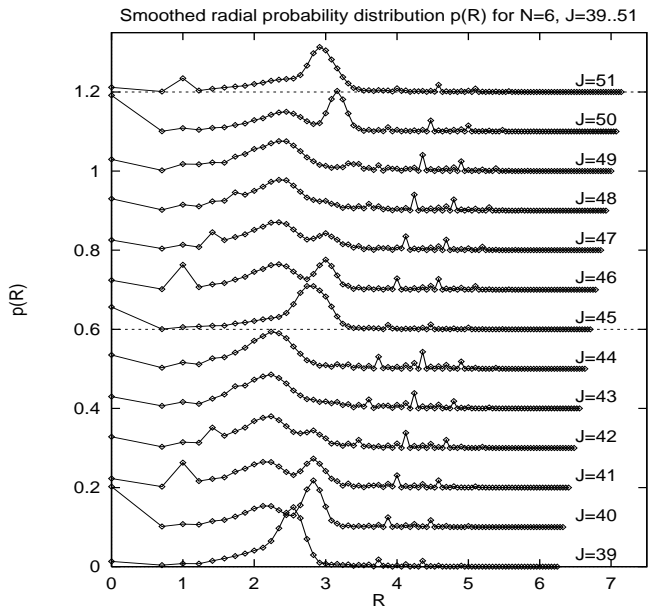} 
    \end{center}
    \caption{Some radial distributions for a quantum dot with $N=6$ electrons
    and total angular momentum $39\leq J\leq 51$. Magic numbers are
    $J=39,45,51$ with $J=45$ being a special magic number. Note that the
    classically expected 5-fold symmetry is only observed for $J=J_{{\rm
    mag}}\pm 1$. This is because the system prefers forming composite
    states which works best for magic numbers. However, the neighbouring
    $J$-values behave similar to atoms with one shell either almost filled
    or filled plus one excess electron.}
\label{fig:R6}
\end{figure}

We can read off the loss of symmetry from the radial distribution
function. At a magic number $J$, we would expect it to have a well defined
peak centered at $R=\sqrt{J}$ for $N\leq 5$, and for $N\geq 6$ it should
show two equally high peaks, one at $R=0$. However, we find something
quite different from that. In the following plot (figure \ref{fig:R6}), 
we compare the radial distributions for a range $39\leq J\leq 51$
around the special magical value $J=45$ for $N=6$. One clearly sees that
at the ordinary magic numbers $J=39,51$ full $\BZ_6$ symmetry is restored
(since there is no peak at $R=0$)! $\BZ_5$ symmetry with one particle in
the center is most prominently exhibited for the neighbouring values
$J=40,50$. This very sensitive dependency on the value of $J$ is a pure
quantum effect contradicting classical expectations. However, the behaviour
at the special magic number $J=45$ is different. There is a certain
probability for a particle to sit in the center, but it is only half of
the probability of the main peak. This tells us that the main peak cannot
come alone from the configuration with a five particle block and one
particle in the center.

One way to interpret this result is to say that a quantum dot behaves
either more as a quantum mechanically system or more as a statistial
ensemble depending on the value of the filling factor $\nu$ defined in
(\ref{eq:nu}). The system is very much like a statistical ensemble and most
resembles\linebreak
\noindent a quantum droplet for $\nu=\frac{1}{2k'+1}$ with $k'\in\BZ_+$
which precisely happens for $J$ being a special magic number. That is the
same condition as for the filling factor for Laughlin-type
fractional quantum Hall states which {\em are\/} described via the
one-component plasma analogy as statistical systems! The system is very
quantum mechanical, and in fact resembles most an atom-like structure,
when $J$ is a {\em non}-special magic number. This does not translate into
a straightforward statement with respect to the filling factor.

The best way to compare the behavior at special magic numbers 
(\ref{eq:smagic}) with the one at ordinary magic numbers (\ref{eq:magic}) 
would be to calculate the 2-point charge density correlation $p(\vec{r},
\vec{r}_0)$. In numerical studies of quantum dots, one usually freezes the
position of one electron at $\vec{r}_0$ and computes the probability of
finding a charge at $\vec{r}$. If $\vec{r}_0$ is chosen such that it
lies in an angular momentum orbital of high probability, one will get
plots which clearly exhibit the symmetry properties of the quantum dot
state. However, to do this we would need the wave function
which we avoided calculating. But the information encoded in 
(\ref{eq:radial}) is sufficient to give us a semi-classical approximation
of the charge distribution. We simply pick the configuration with the
lowest energy and put us into its ``rest frame''. By this we mean the
following: The minimal energy configuration has a certain block
decomposition, and for this\linebreak
\vspace{-0.5cm}\begin{figure}[h]
    \begin{center}
    \leavevmode\epsfxsize=44mm\epsfysize=44mm\epsfbox{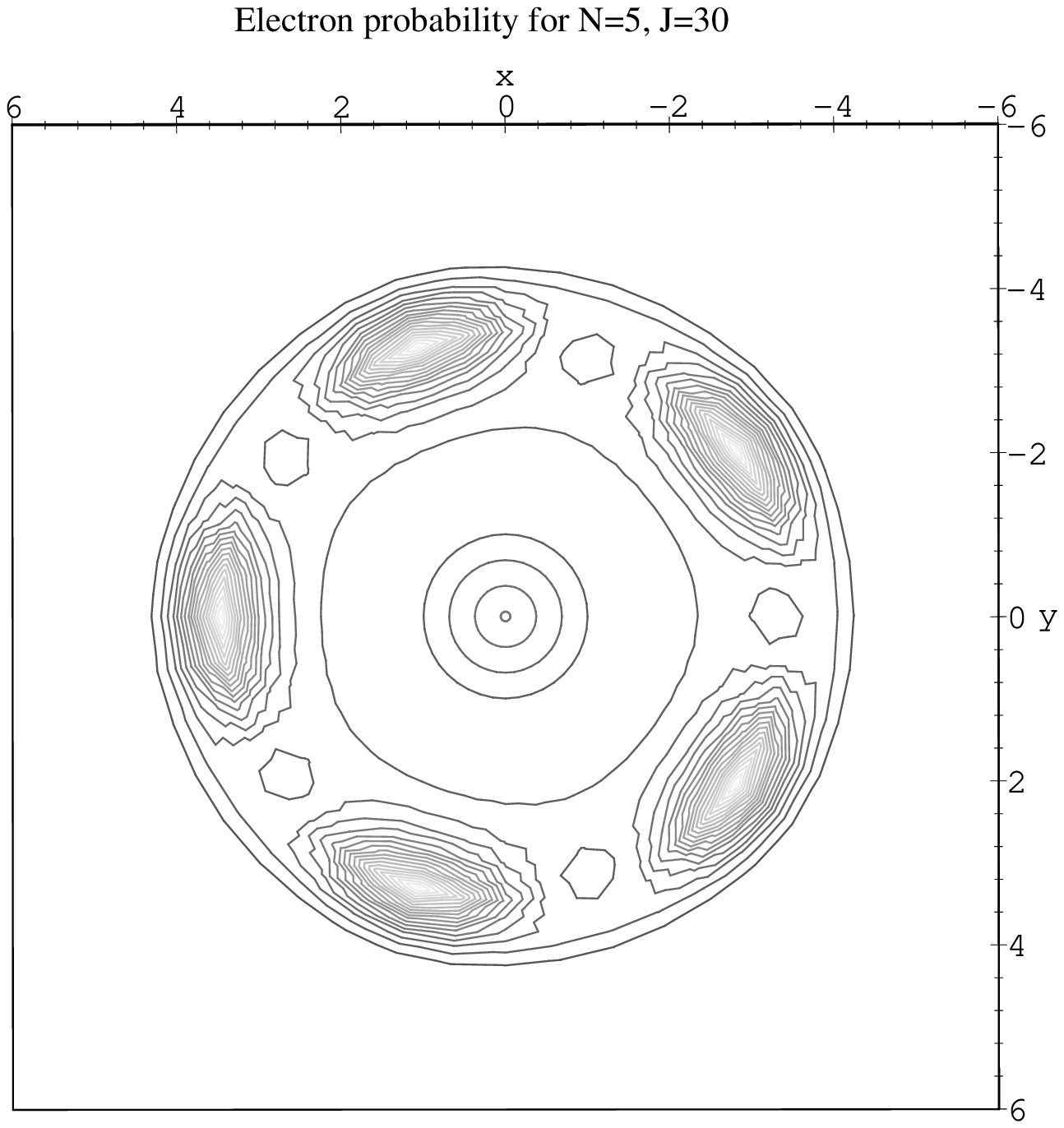%
    }\epsfxsize=44mm\epsfysize=44mm\epsfbox{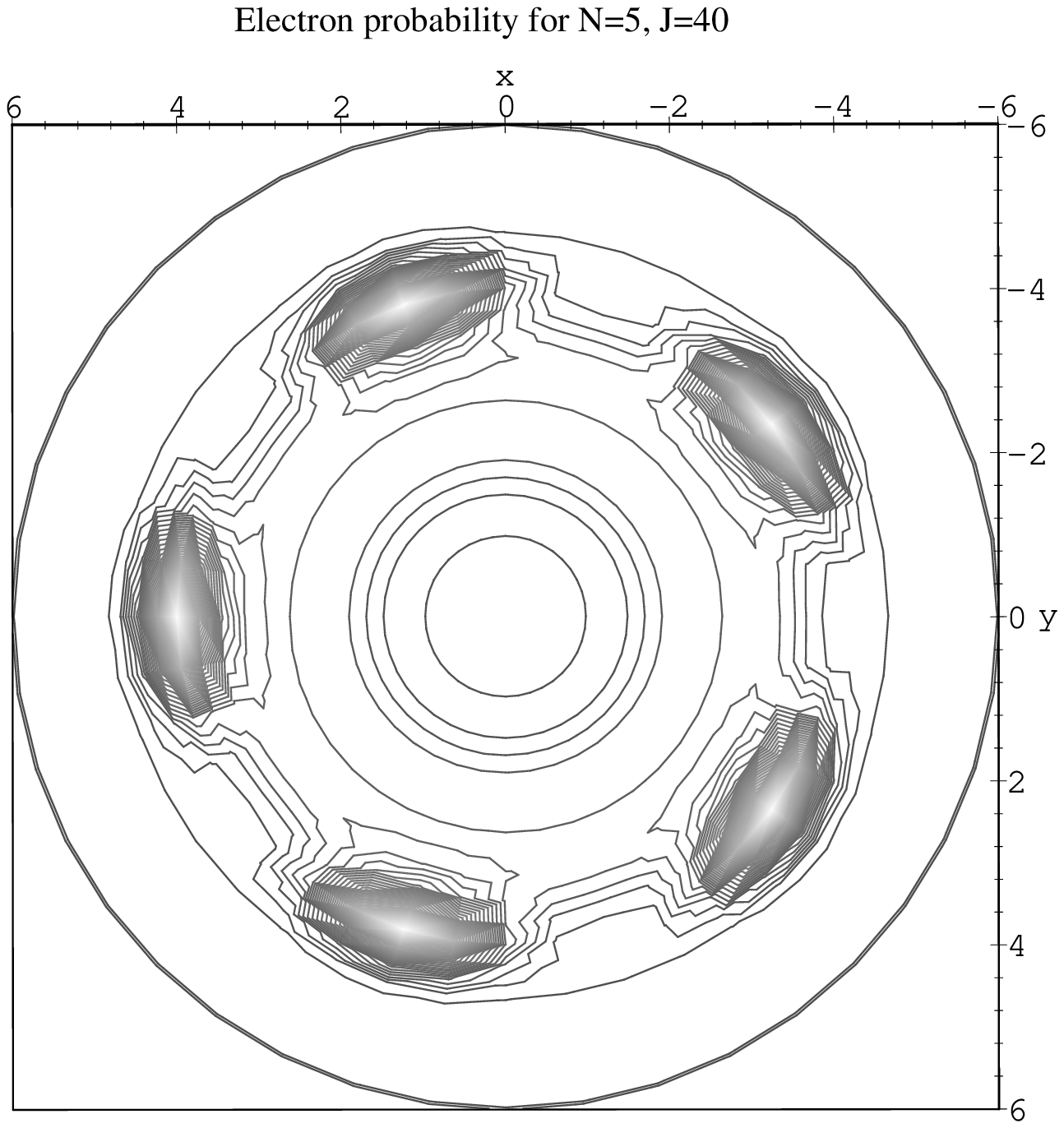}\\
    \leavevmode\epsfxsize=44mm\epsfysize=44mm\epsfbox{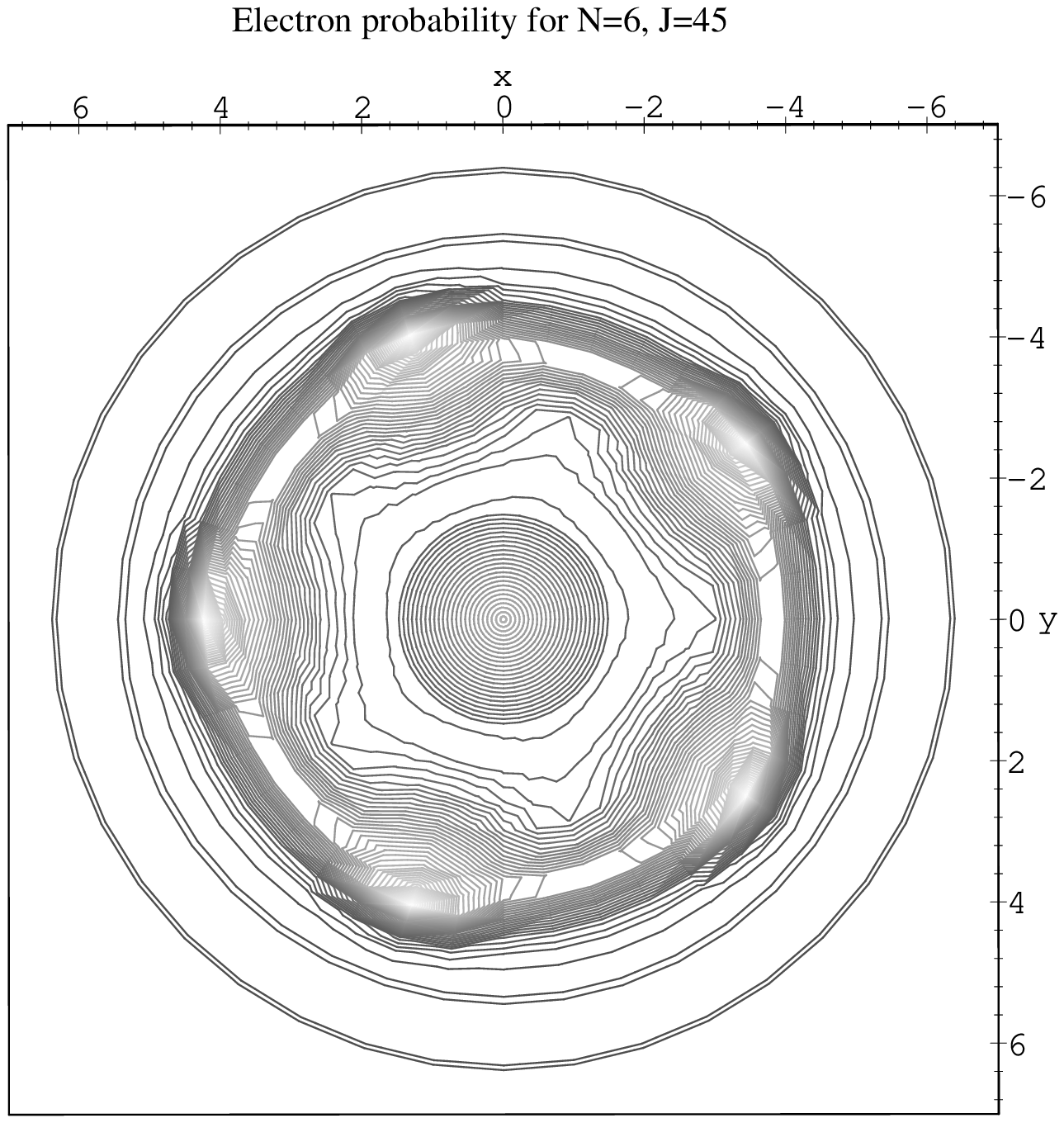%
    }\epsfxsize=44mm\epsfysize=44mm\epsfbox{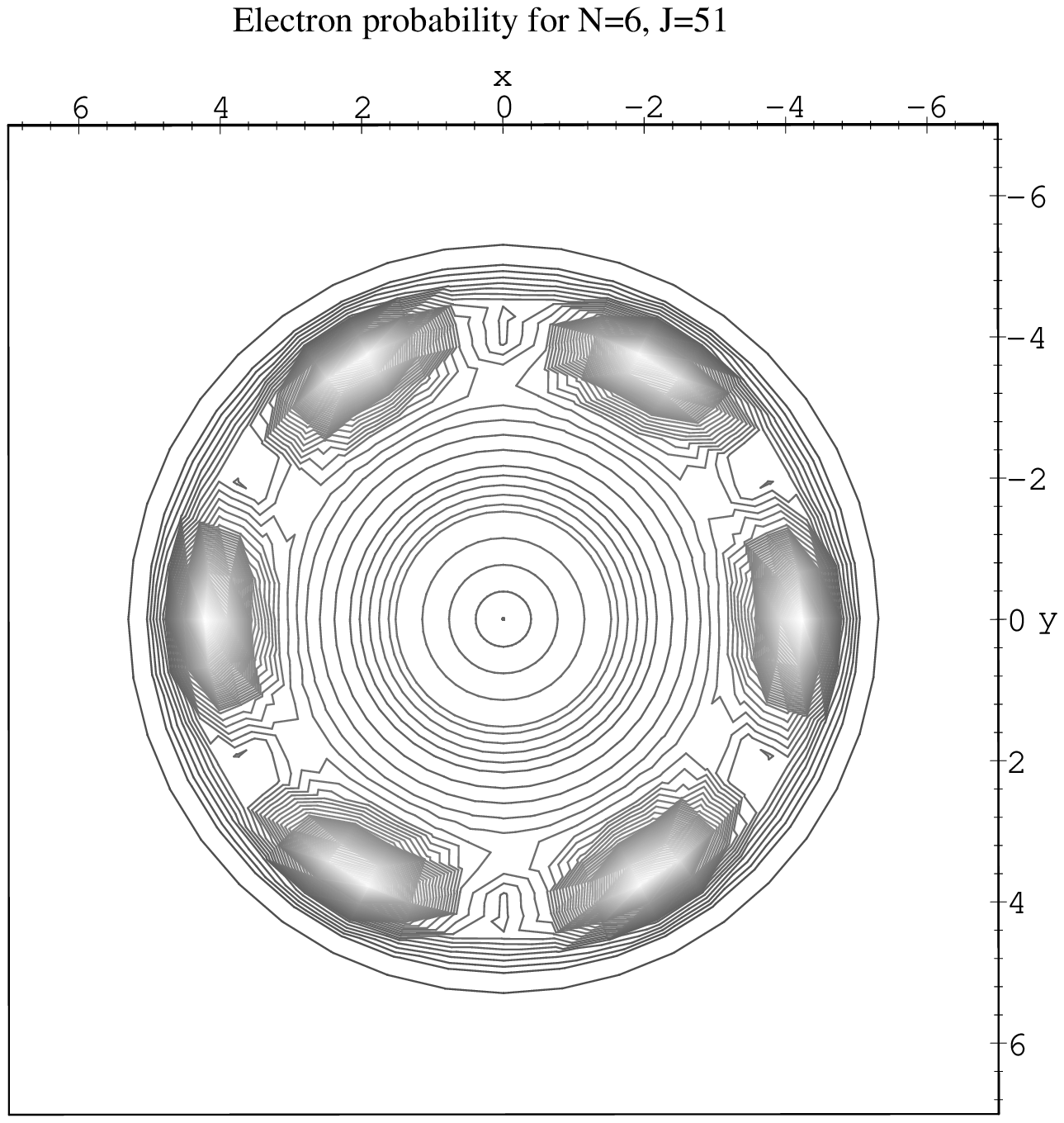}
    \end{center}
    \caption{Comparsion of approximated charge distributions (in the
    rest frame of the minimal energy configuration) for $N=5$ (top) and
    $N=6$ (bottom). Left are special magic numbers corresponding to
    $\nu=1/3$, right are ordinary magic numbers. Note the difference
    for $N=5$ and $N=6$, the latter showing much stronger quantum droplet
    behavior at a special magic number.}
\label{fig:P56}
\end{figure}
\noindent decomposition we localize the electrons
on the edges of the appropriate $n$-gons, smeared out according to
the binominal distribution taking into account the quantum mechanical 
width of the $n$-gons. For $N\leq 5$ the minimal configuration is 
simply one $N$-gon, for $6\leq N\leq 8$ it is a $(N-1)$-gon with an
occupied center. For $N\geq 9$ we have a shell like structure where
the relative orientation of the electrons in one shell to the ones
in another would have to be chosen according to the classical
solution. All other configurations are considered to be radially
symmetric. In this way, superimposing these approximations of
charge distributions, we obtain a measure of how much
the minimal energy configuration sticks out of the other
ones, and hence a measure of how
much the particular symmetry of the minimal energy configurations
determines the symmetry of the full charge distribution. When plotting
charge distributions in this manner, it is clear that the integrated
radial probability for a block within a configuration 
(first equation in \ref{eq:radial}) has to be weighted not only by
the binominal distribution, but also by $1/2\pi r$
where the radius $r$ varies over the width where the block is localized. 
Moreover, such a charge distribution is naturally time averaged. 
To be more precise, we approximate the probability distributions as
follows: A block $(n,\jb)$ with probability $p$, if it does not have the 
dominant symmetry of the minimal energy configuration, contributes
\be
  \tilde{p}(j,\phi) = p\,{\cal D}_{n/2}(j-\jb)
  \frac{1}{2\pi\sqrt{j+1}}\,,
\ee
where ${\cal D}_{\sigma}(x)$ denotes a symmetric distribution centered 
around zero with width $\sigma$, in our case a normalized binominal 
distribution such as ${\cal D}_{\sigma}(x) = 
2^{-2\sigma}\Gamma(2\sigma+1)/\Gamma(\sigma+1-x)\Gamma(\sigma+1+x)$. Since
this is independent of $\phi$, it is radially symmetric. Of course, as
mentioned above, we could equally well have chosen another distribution
such as a Gaussian one without affecting the final results very much.
In the case where the block has the dominant symmetry of the
minimal energy configuration, we use instead 
\bea
  \tilde{p}(j,\phi) &=& p\sum_{k=0}^{n-1}{\cal D}_{n/2}\left(\sqrt{j^2+\jb^2
  - 2j\jb\cos(2\pi(k/n-\phi))}\right)\nonumber\\
  &\times& \frac{1}{2\pi\sqrt{|j-\jb|+1}}\,.
\eea
Our approximation differs from the approach where one
electron is frozen on a likely position only in the fact that we freeze 
the position of one suitable composite particle state. Therefore, our
scheme is essentially only the consequent translation of computing
pair correlations
$p(\vec{r},\vec{r}_0)$ in a single particle basis to our basis of
composite particle states. As figure \ref{fig:P56} clearly demonstrates,
full $N$-gon symmetry holds even for $N\geq 6$ for ordinary magic numbers,
while symmetry loss occurs for special magic numbers and $N\geq 6$.

\section{Conclusion}

Although space permits only to present these few examples,
we have performed calculations of quantum dot states for $2\leq N\leq 9$
and with total angular momentum up to $J=126$. We can confirm the
occurence of special magic numbers for $N=6,7,8$ corresponding to
filling factors $\nu=1/3,1/5,1/7$, surpassing the range of $N,J$
accessible to exact diagonalization methods. We are also able to
see that for $N=9$ more complicated shell-like structures appear, but
that magic numbers still lead to an enhancement of symmetry, and
special magic numbers to a weakening of symmetry. The precise results
will appear elsewhere.

We proposed a new and
extremely simple way to understand the qualitative features of quantum
dots by mixing classical and quantum mechanical points of view. We think
that the results justify our approximation scheme, where we chose a
basis of composite particle states for which only the classical
repulsive interaction is taken into account. Quantum mechanics entered
the game mainly through the quantization of angular momentum which
made the computation of the partition function a finite (and in fact
quite simple and fast) enterprise. As a further simplification, we
made use of universality and
replaced the Coulomb interaction by a $V=|\vec{r}|^{-2}$ potential.
This allowed us to perform all computations with integer arithmetics
alone. Once one accepts our special choice of a basis of states, all
further computations can hence be done exactly within the scheme.
Resorting to the physical Coulomb potential makes computations
slighly more complicated but does not, as we have checked, change
the qualitative features of the results. The main influence of varying
$\gamma$ in $V=|\vec{r}|^{-\gamma}$ is that the relative radial positions of
blocks are shifted somewhat. More precise statements about universality
can be found in \cite{DMV98}.

Of course, our approach is over-simplifying and cannot capture finer
details of quantum dots. There is one particular point of possible
criticism: We assumed that paricles within a block would distribute
themselves equidistantly on a circle. Classically this is not true
in general if the configuration consists of more than one non-trivial
block. However, our central hypothesis suggests that
quantum mechanically each block should be treated as a smooth
delocalized ring shaped charge distribution, since we assume that each
block forms a collective state. Indeed, when
calculating the potential energy between two blocks via (\ref{eq:E2n-gons}),
we assume precisely that. On the other hand, when we calculate the
self-energy of a block, we resort to a classical picture of $n$
point-sources distributed equidistant on a circle, see (\ref{eq:En-gon}).
It is here, where we are in danger of a systematic error, slightly 
underestimating the self-energy in the presence of another block (which
may force the electrons to move to a non-equidistant distribution).
However, if our central hypothesis is true, it depends on the strength
of the coupling of the electrons within a block how much the resulting
collective state is influenced by the presence of another block.
In our simplistic picture, we assume that such corrections are
neglegible and do not change the qualitative patterns. This is
plausible since spin-spin interactions among completely polarized
electrons lead to a further repulsive interaction which locally, within
a block, might dominate effects coming from the electro-static interaction
with other blocks.

Another possible point of criticism is our normalization condition
(\ref{eq:norm}) which assumes ortho-normality of our basis of states.
This is of course not entirely true. However, figures 1 and 2 clearly
show that the resulting temperature estimates are not too far off what
we would expect, namely $\vev{E}\sim NkT$ for $N$ particles. In fact,
one can easily replace (\ref{eq:norm}) by the condition
${\cal Z}(\beta) = C\beta^{-N}$ with $C$ chosen such that for the
minimal possible case, $J=N(N-1)/2$, ${\cal Z}(\beta) = 1$. Solving this
constraint is slightly more complicated, but the other results differ
not very much from the ones obtained with our idealistic\linebreak 
\vspace{-0.5cm}\begin{figure}[h]
    \begin{center}
    \leavevmode\epsfxsize=44mm\epsfysize=44mm\epsfbox{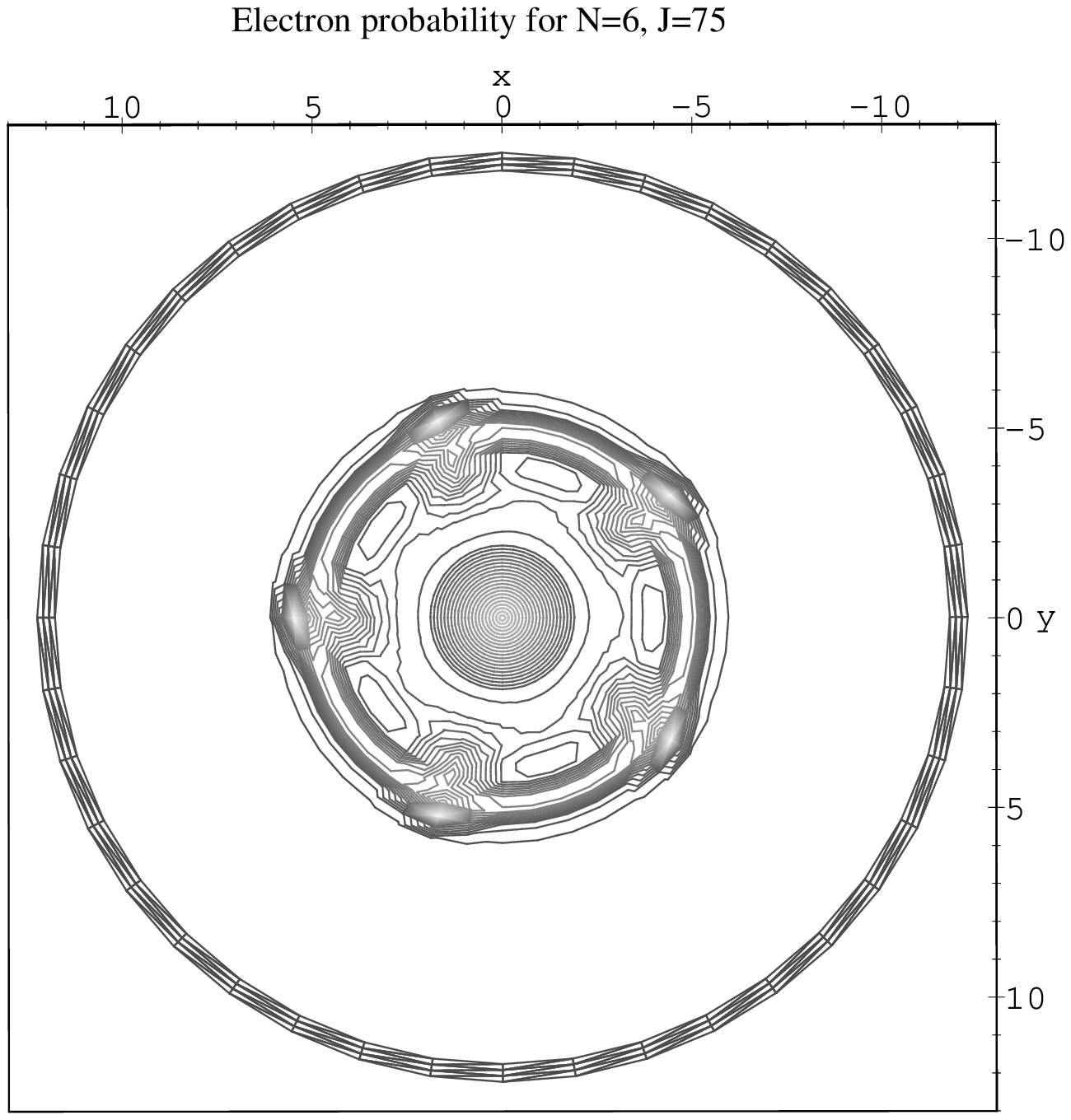%
    }\epsfxsize=44mm\epsfysize=44mm\epsfbox{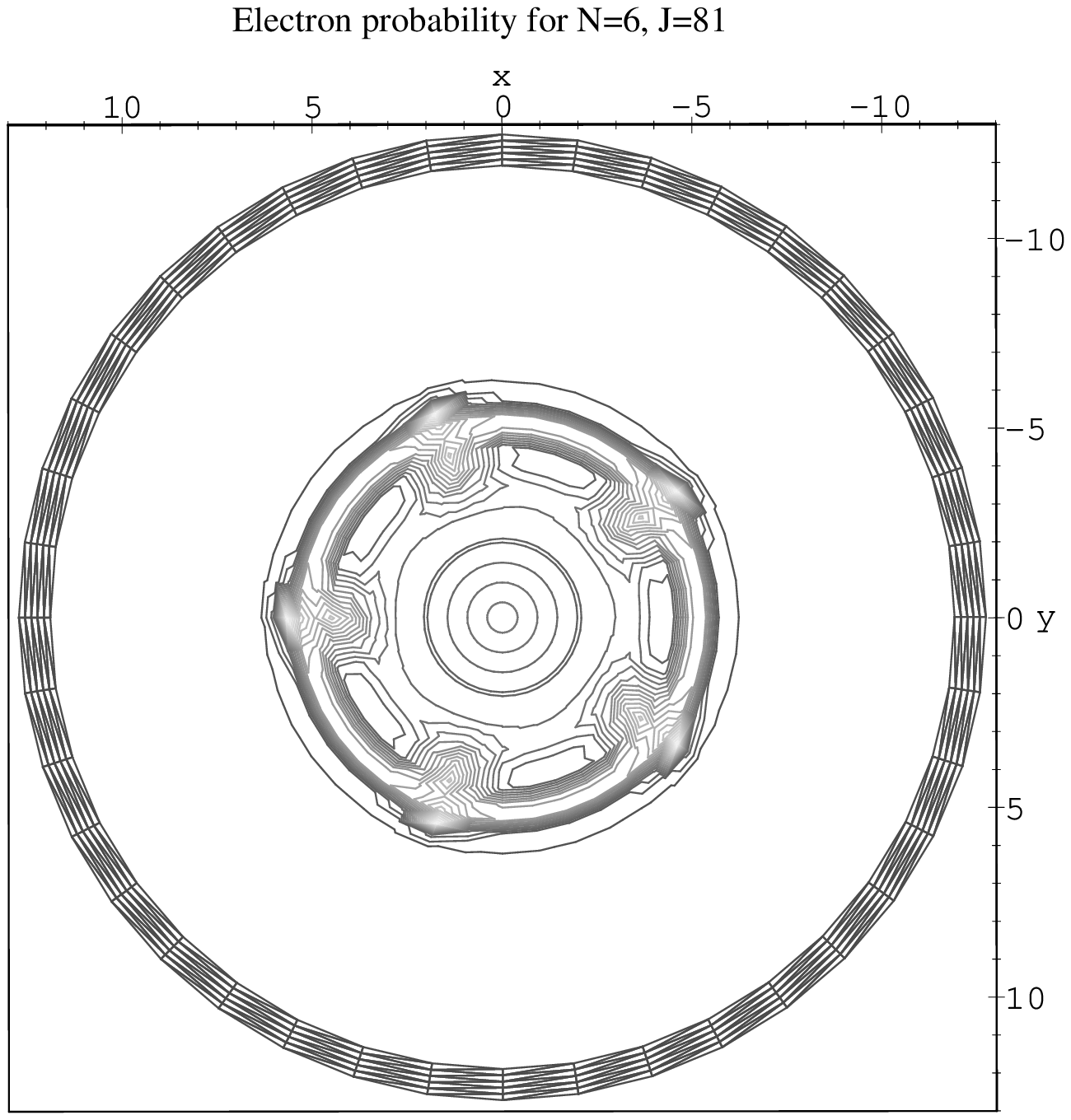}\\
    \leavevmode\epsfxsize=44mm\epsfysize=44mm\epsfbox{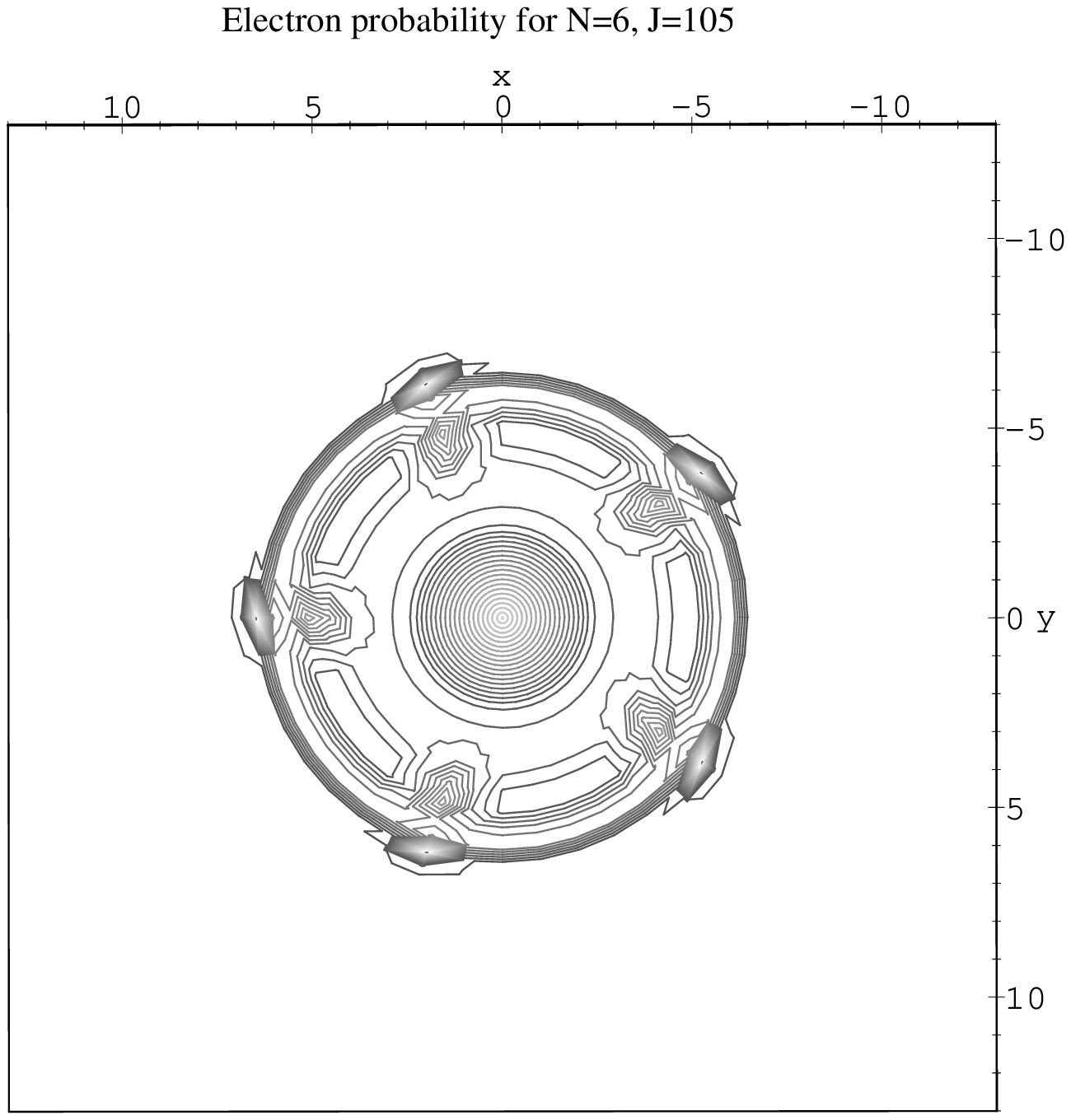%
    }\epsfxsize=44mm\epsfysize=44mm\epsfbox{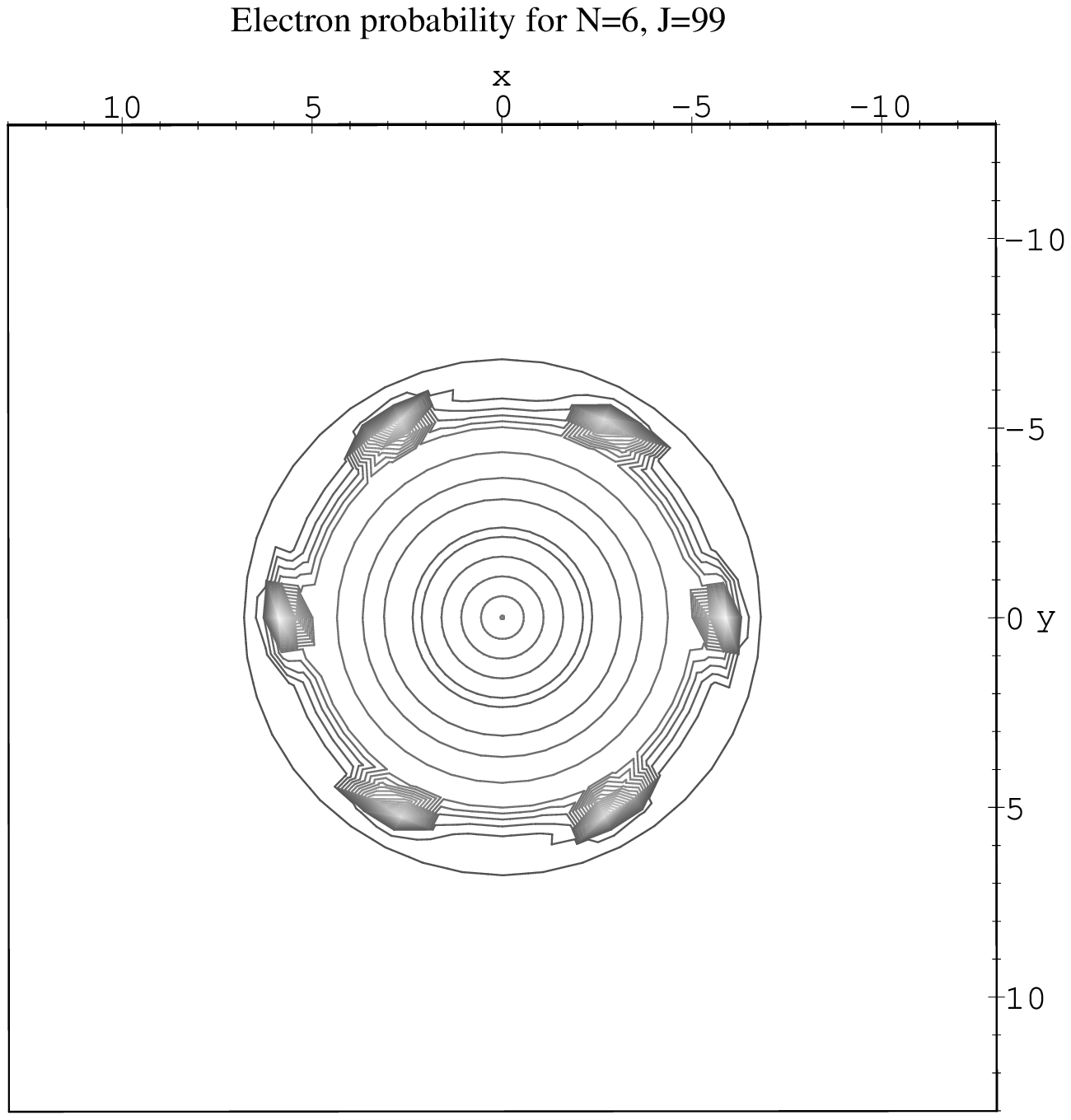}
    \end{center}
    \caption{Comparsion of approximated charge distributions (in the
    rest frame of the minimal energy configuration) for $N=6$. Left are
    the distributions for filling factors $\nu=1/5$ (top) and $\nu=1/7$
    (bottom) with $J=75$ and $J=105$ respectively. 
    This is compared with the distributions for non-special
    magic numbers close to the special magic ones on the right. Note that
    no full restoration of symmetry occurs for $J=81$ (top), where the $j=1$
    orbital can be seen to be occupied, but where the lowest-energy 
    configuration barely sticks out of the more likely fully
    symmetric configuration. In case of $J=99$ (bottom), full symmetry
    restoration is still observed. See main text for details about which
    magic numbers fail to completely restore maximal symmetry.}
\label{fig:largeJ}
\end{figure}
\noindent approach. In
particular, the radial distributions just become a bit more smeared out, but 
retain their qualitative structure.

Our results bring us to the conclusion that our hypothesis is a good
one, i.e.\ that it predicts the correct qualitative behavior expected
from quantum dots. It would hence be very interesting to derive this
hypothesis from a microscopical treatment of quantum dots. It is well
known that spin-spin and spin-orbit interactions are important in the
theoretical understanding of electron orbitals of atoms. Quantum dots
are often described as ``artificial atoms'' \cite{atom}, 
and it would only further
justify this labeling, if they indeed shared the importance of
spin-interaction effects with their name-inducing natural relatives.
Also, it would be interesting to relax our condition of completely
polarized electrons in order to study quantum dots where angular momentum
orbits can be occupied by upto two electrons. Finally, a more realistic
normalization condition for the partition function, as mentioned above,
might be desirable.
But these investigations will be left for future work.

\vspace{-0.2cm}\begin{figure}[h]
    \begin{center}
    \leavevmode\epsfxsize=44mm\epsfysize=44mm\epsfbox{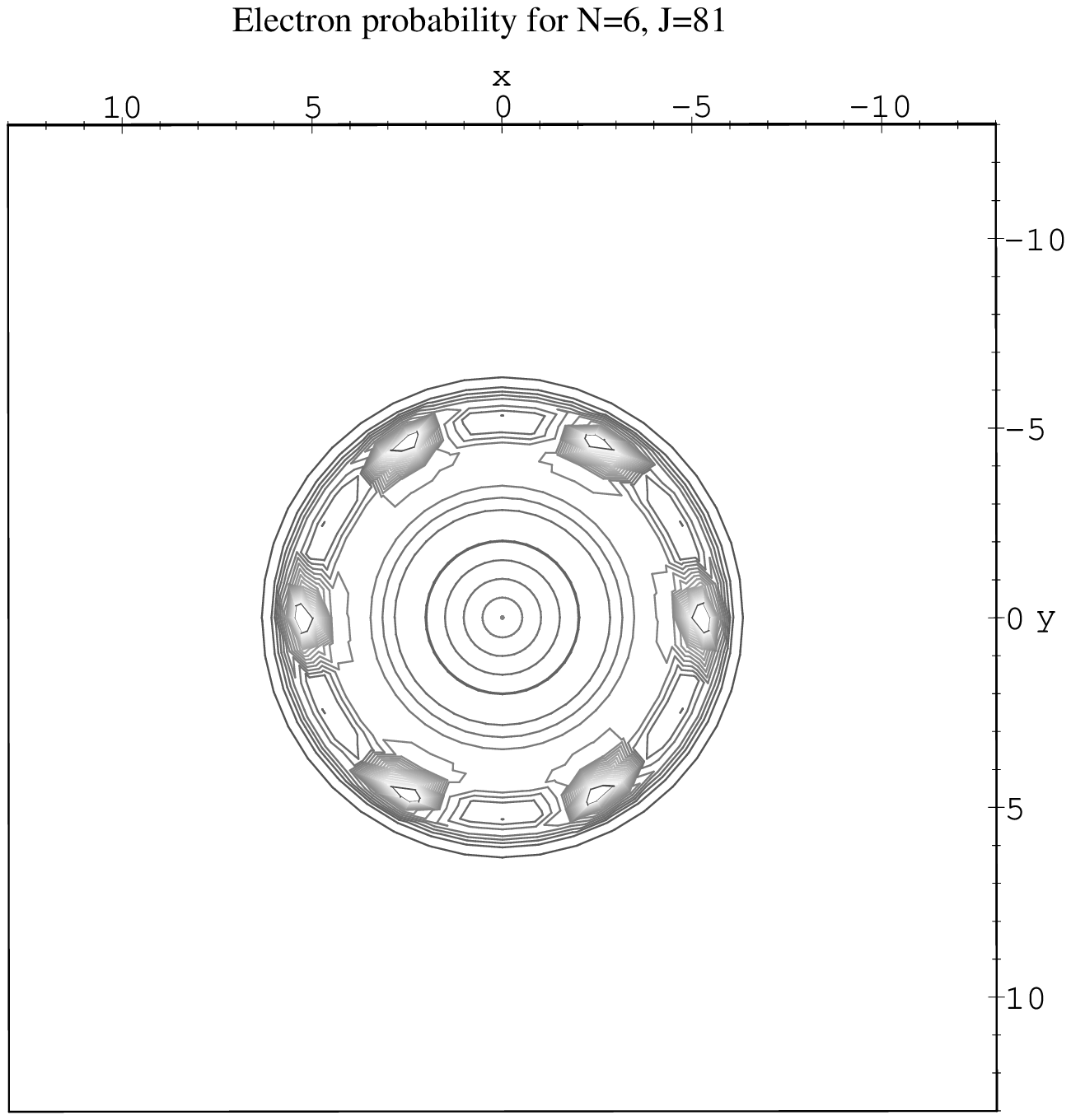%
    }\epsfxsize=44mm\epsfysize=44mm\epsfbox{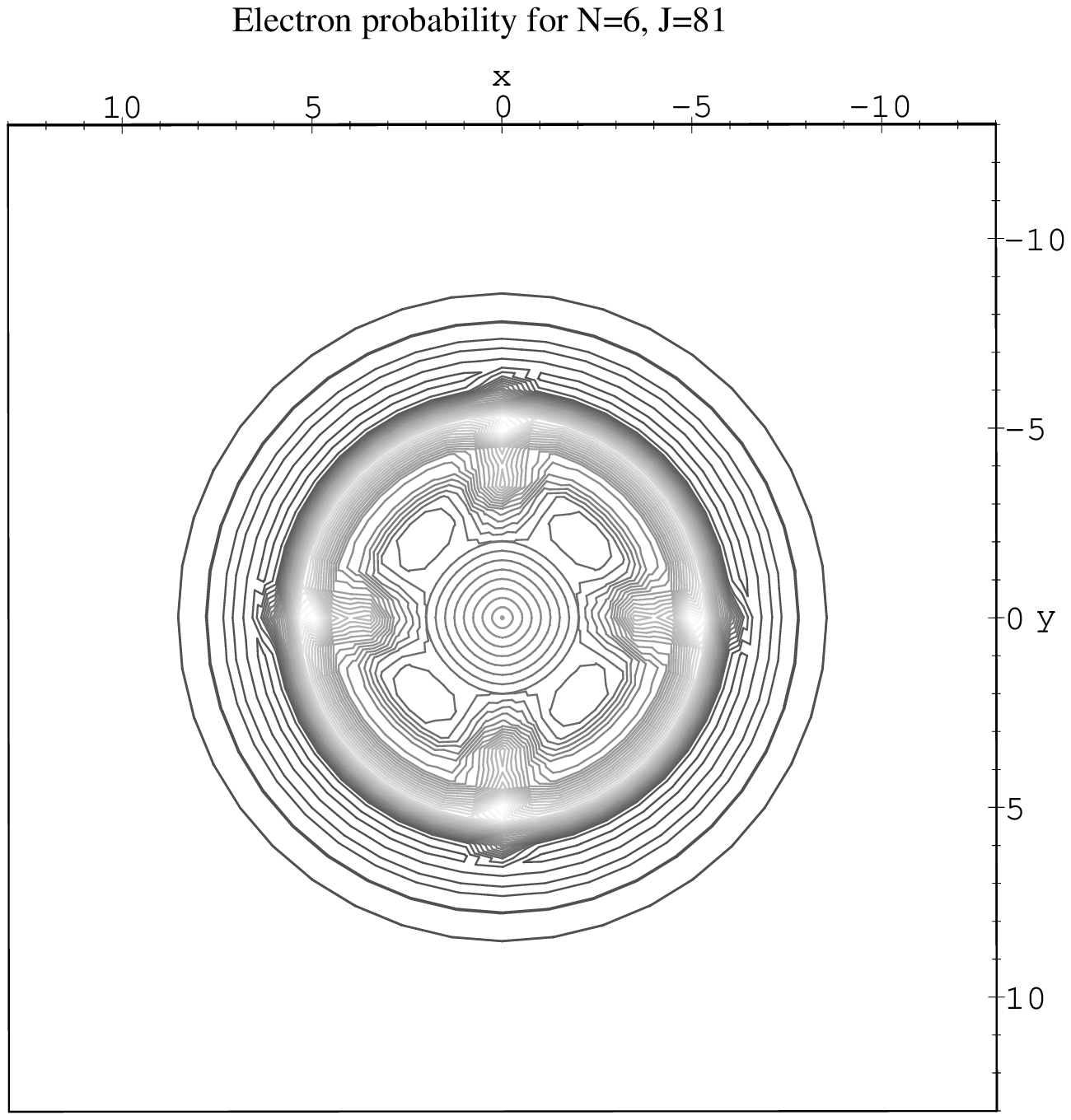}\\
    \leavevmode\epsfxsize=44mm\epsfysize=44mm\epsfbox{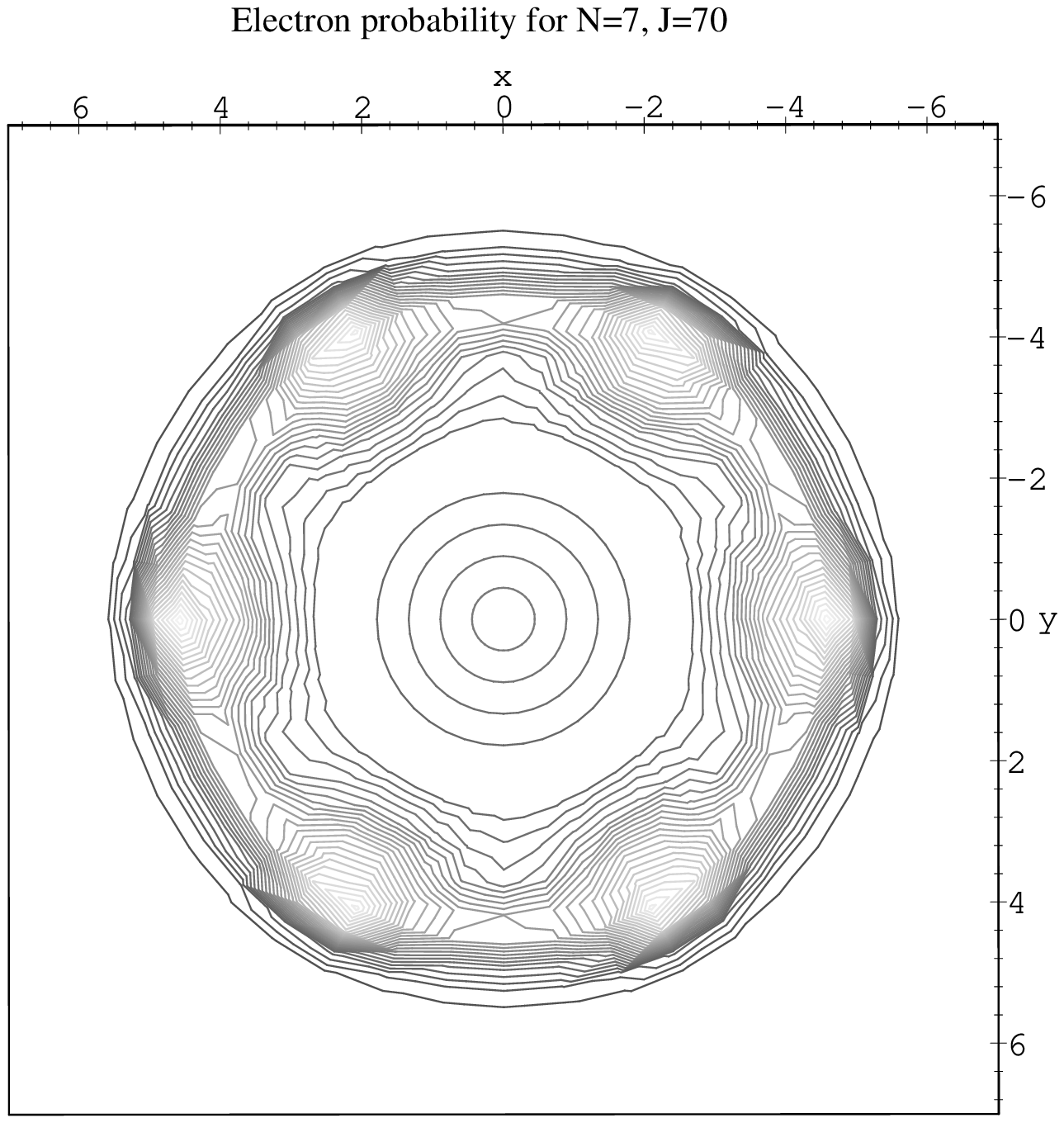%
    }\epsfxsize=44mm\epsfysize=44mm\epsfbox{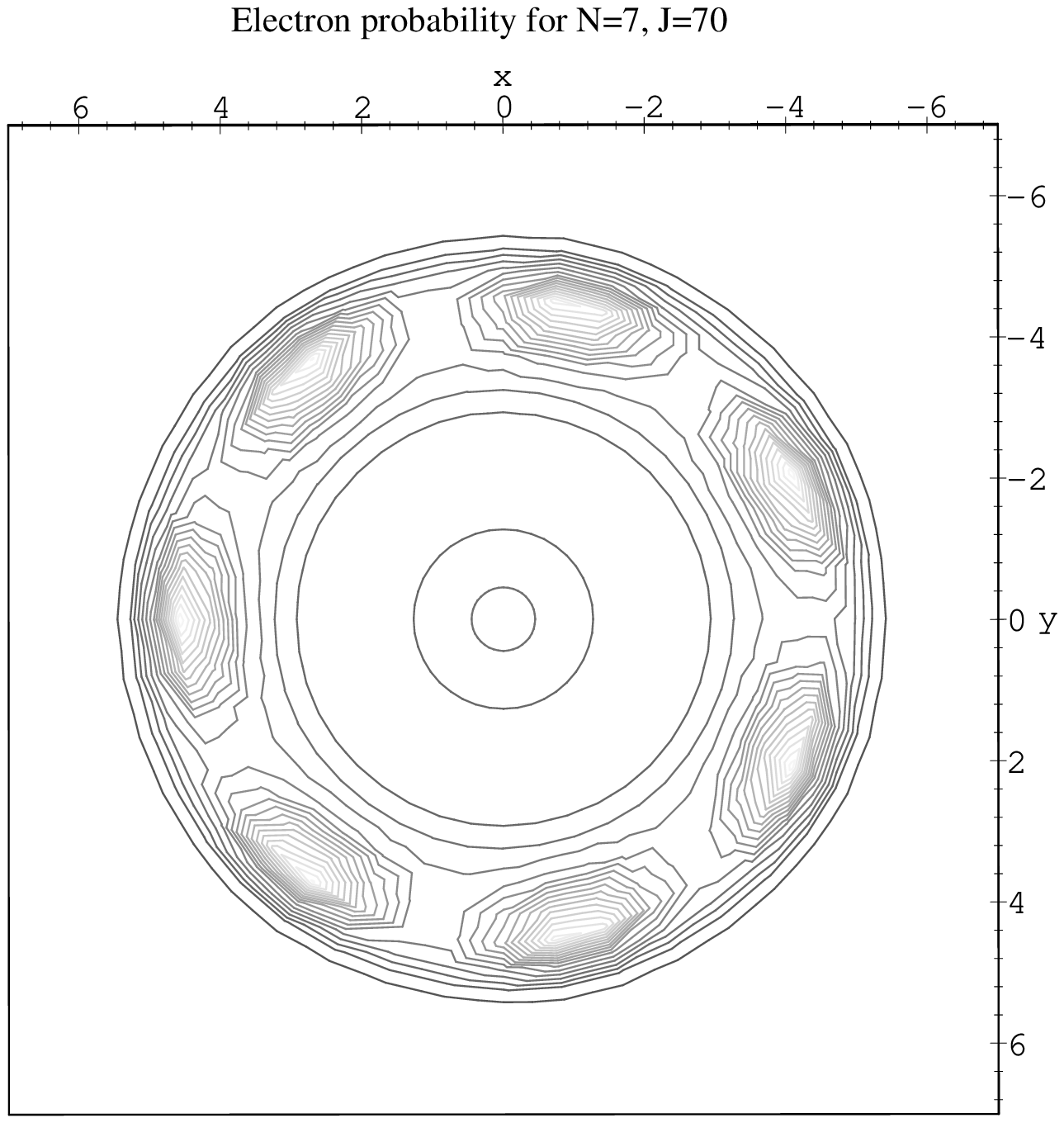}
    \end{center}
    \vspace{-0.2cm}\caption{Comparsion of approximated charge 
    distributions for different
    rest frames. Top row shows the quantum dot for $N=6$ and $J=81$
    in the rest frames for full 6-fold symmetry (left) and
    for 4-fold symmetry (right), to be compared with the plot for 
    5-fold symmetry in figure 5 (top right). The fully
    symmetric configuration is still the dominant one. Bottom row shows
    a similar comparsion for $N=7$ and $J=70$ (the first magic number
    after the special magic number $J=63$ for $\nu=1/3$). The difference
    between the lowest energy rest frame (left) and the highest symmetry
    rest frame (right) is very small, but the latter is the dominant
    one. Hence, symmetry restoration still takes place, but is much 
    weaker expressed than for other non-special magic numbers.}
\label{fig:restframes}
\end{figure}
\section{Predictions}

So far, we have successfully reproduced results which were already
achieved by other methods, mainly by exact diagonalization techniques.
Since our method is simple and fast, we can probe higher particle
numbers and total angular momenta than the ones accessible to numerical
methods. We will present here some calculations for $N=6$ with much
larger total angular momentum as can be found in the literature
(see figure \ref{fig:largeJ}).

In particular, we confirm that symmetry loss -- as expected -- occurs for
higher special magic numbers corresponding to fillings $\nu=1/5$ and
$\nu=1/7$. On the other hand, full $\BZ_6$ symmetry is still restored
for {\em most but not all\/} non-special magic numbers! This is a new
phenomenon, and also a nice demonstration of mesoscopic physics. 
Most non-special
magic numbers do not contain states of the form $[(1,\jb_1),(N-1,\jb_2)]$
with energies comparable to the (for magic numbers always existing)
configuration $[(N,\jb=J/N)]$. The reason is that for most magic numbers
there is no choice with very small $\jb_1$ which would favor a smaller
energy than the energy of the $[(N,\jb=J/N)]$ configuration. 
One of the properties which make special magic numbers special is that 
at these values of the total angular momentum the configuration 
$[(1,0),(N-1,J/(N-1))]$ becomes possible. 

However, if the total angular momentum $J$ is sufficiently large,
configurations such as $[(1,1),(N-1,(J-1)/(N-1))]$ may have energies smaller
than the energy of the most symmetric solution. In the case of $N=6$ this
happens the first time for $J=81$, where the above given configuration 
consisting of a block with 5 particles around a single particle orbiting
the center in the $j=1$ orbital has indeed a slightly lower energy than the
configuration with a full block out of 6 particles, although the former
state has a much smaller probability than the latter. This is simply due to
the fact that for $J\gg 1$, the difference between the energies of the
configurations $[(1,0),(N-1,J/(N-1))]$ and $[(1,1),(N-1,(J-1)/(N-1))]$ 
becomes arbitrarily small. Put differently, for $J\gg 1$ the system
behaves more and more classical, but the transition from the quantum
realm to the classical world is not continuous with increasing $J$.
For example, full $\BZ_6$ symmetry is restored for the magic number
$J=99$. More precisely, the first non-special magic numbers (\ref{eq:magic})
where symmetry restoration does not take place are $J(N,k'N)$ for $k'\geq 
k'_0(N)$. We have for example $k'_0(6)=2$ and $k'_0(7)=1$.
These happen to be the non-special magic numbers which follow next to
special magic numbers. For even higher $J$ we might expect that more and
more magic numbers fail to restore the full symmetry of the system.

Because the minimal energy configuration does not possess the maximal
possible symmetry, it may have a smaller probability than the maximal
symmetric configuration due to different multiplicities. In fact, this
is the case when the magic numbers are non-special. When the magic
number is special, the minimal energy configuration is the dominant
one, although the dominance is not strongly expressed -- which is precisely
the reason which leads to the
droplet like probability distributions of such quantum dot states. 
It follows, that symmetry restoration should still work to some degree
for the above discussed non-special magic numbers.
We demonstrate this in figure \ref{fig:restframes}, where we compare
probability distributions for different rest frames, namely the rest
frame of the minimal energy configuration and the one for the maximal
symmetry configuration. One can infer from these plots that symmetry
restoration does take place for the non-special magic numbers $J(N,k'N)$, 
but of a somewhat lesser degree than for other non-special magic numbers.

Therefore, we have demonstrated that our method is capable of probing 
unknown regions in ``quantum dot space'' showing and predicting new patterns. 
We believe that our method is a useful tool to explore in more detail
how quantum mechanics and classical physics are interlinked in mesoscopic
systems. 

\section*{Acknowledgment} 

It is a pleasure to thank
Sarben Sarkar and Charles Creffield who got me interested in the
fascinating field of quantum dots in the first place, and with whom
I had many stimulating discussions. I would also like to thank
Gerard Watts, Mathias Pillin and Werner Nahm
for valuable discussions and comments.

\end{multicols}
\end{document}